\documentclass[preprint,           
               noshowpacs,          
               preprintnumbers,     
               aps,                 
               prd,                 
               a4paper,             
               superscriptaddress,  
               tightenlines,        
               floats,floatfix,      
               nofootinbib         
               ]{revtex4}

\usepackage{amsmath, amsfonts, amsthm, amssymb, graphicx}

\def\be{\begin{equation}}
\def\ee{\end{equation}}
\def\ba{\begin{eqnarray}}
\def\ea{\end{eqnarray}}
\newcommand{\nn}{\nonumber\\}
\newcommand{\ud}{\mathrm{d}}
\def \pd {\partial}
\def \bfx {{\bf x}}

\begin{document}

\title{Goldstone's Theorem and Hamiltonian of Multi-galileon Modified Gravity}

\author{Shuang-Yong Zhou} 
\email[]{ppxsyz@nottingham.ac.uk}

\affiliation{School of Physics and Astronomy, 
University of Nottingham, Nottingham NG7 2RD, UK} 

\date{\today}

\begin{abstract}

The galileon model was recently proposed to locally describe a class of modified gravity theories, including the braneworld DGP model. We discuss spontaneous symmetry breaking of the self-accelerating branch in a multi-galileon theory with internal global symmetries. We show a modified version of Goldstone's theorem is applicable to the symmetry breaking pattern and discuss its implications. We also derive the Hamiltonian of a general multi-galileon theory and discuss its implications.

\end{abstract}


\maketitle

\section{Introduction}

The DGP model~\cite{DGP,sa} is a 5 dimensional braneworld theory that non-trivially modifies General Relativity (GR) in the infrared. Nevertheless, at sub-crossover (sub-Hubble Length) scales many of its properties can be captured by a 4D (boundary) effective theory~\cite{dgppi,dgppi1}. This effective theory amounts to GR coupled to a scalar field $\pi$ whose equation of motion has only second derivatives and is invariant under the Galilean shift $\pi \to \pi+a_{\mu}x^{\mu}+b$, $a_{\mu}$ and $b$ being constant. This scalar is related to the bending of the DGP brane in the bulk and has been termed as \emph{galileon}~\cite{galori}.

As ghost instability has been identified on the phenomenologically interesting self-accelerating branch of the DGP model~\cite{saghosts}, which can also be easily seen in the local galileon approximation~\cite{dgppi,dgppi1}, attempts have been made to generalize the DGP galileon description to produce a healthy modified gravity theory~\cite{galori,covgal,covgal1,dbigal,dbigal1,pforms,psz,psz1,psz2,htw,ahkt}. In~\cite{galori}, the authors wrote down the most general single galileon Lagrangian. Remarkably, there are only $d+1$ possible galileon terms in $d$ dimensional spacetime, and ghost free self-accelerating background solutions have been shown to exist in a generalized galileon theory. However, a few phenomenologically challenging problems have also been identified in the single galileon theory, such as Cherenkov-like radiation in the solar system, superluminal propagation far away from a matter sauce and very low strong coupling scales~\cite{galori}. It turns out that these problems can be avoided by adding another galileon (in a bi-galileon theory), meaning the theory space of the single galileon model is actually too small~\cite{psz1}. A local bi-galileon description is also what one might expect from co-dimension 2 braneworld models~\cite{psz,htw,cod2}, as there are generally two brane bending directions.

One would want to generalize the galileon description to have even more degrees of freedom~\cite{psz2,pforms,htw,ahkt}. To avoid a proliferation of possible terms in the theory, we can impose internal (global) symmetries within the multiple galileons, so that the multiple galileons form some representation of a group~\cite{psz2}, $\pi=(\pi_1,...,\pi_N)$. That is, the multi-galileon Lagrangian is imposed to be invariant under the internal transformation
\be
\pi_i\to {\cal R}_i^{~j}\pi_j,
\ee
where ${\cal R}_i^{~j}$ is the representation matrix of a certain group and summation over repeated group indices is implied. Notice that the internal symmetry could originate from braneworld scenarios, as has been identified for the $SO(N)$ fundamental representation~\cite{psz2,htw}. For other interesting field theoretical and cosmological implications of the galileon theory, see~\cite{galsim,cos1,cos4,cos2,cos3}.

In~\cite{psz2}, we wrote down all possible multi-galileon terms that are consistent with the fundamental and adjoint representations of $SO(N)$ and $SU(N)$, and looked for soliton solutions in multi-galileon theories; We did not consider coupling the symmetric multi-galileon to gravity. In this paper, we will put the symmetric multi-galileon in the context of modified gravity. In Section~\ref{coupgravity}, we will venture a tentative coupling, but we want to emphasize that the main results of this paper are insensitive to this explicit coupling. In Section~\ref{goldstonet}, we discuss the spontaneous symmetry breaking phenomenon of the symmetric multi-galileon theory on a self-accelerating background. Starting from an example, we build up a new version of Goldstone's theorem in symmetric multi-galileon theories that for every broken continuous symmetry a canonical kinetic degree of freedom is lost. In Section~\ref{hamil}, we derive the Hamiltonian formulation of a general multi-galileon theory (with or without internal symmetry) and find it is not bounded below. We speculate whether this might be overcome in more complete theories.

\section{Multi-galileon Modified Gravity} \label{coupgravity}

In the original galileon model~\cite{galori}, the galileon is coupled to graviton mainly via the kinetic mixing
\be
h_{\mu\nu}=\tilde{h}_{\mu\nu}+2\pi \eta_{\mu\nu},
\ee
where $h_{\mu\nu}$ and $\tilde{h}_{\mu\nu}$ are Jordan and Einstein frame (perturbative) metrics; $\pi$'s contribution to the energy momentum tensor, or, its \emph{direct} influence to the geometry is negligible. So in a sense the galileon modified gravity is a ``genuine'' infrared modification of General Relativity, differing from models such as quintessence~\cite{quint}, which has significant contribution to the energy momentum tensor. In this paper, we stick to this paradigm and tentatively propose the multi-galileon's coupling to gravity as
\begin{align}\label{coupling}
S=\int \ud^4 x   \left[ -\frac{M_{P}^2}{4}\tilde h^{\mu\nu}{\cal E} \tilde h_{\mu\nu}+\frac{1}{2} \tilde h_{\mu\nu} T^{\mu\nu}
  +(\pi_1+...+\pi_N) T +{\cal L}_{\pi} \right]  ,
\end{align}
where $T\equiv \eta_{\mu\nu}T^{\mu\nu}$ and ${\cal L}_{\pi}$ is the multi-galileon Lagrangian. For a general multi-galileon theory without internal symmetries, we might want to redefine $\pi'_1=\pi_1+...+\pi_N$ to simplify the coupling, while keep the structure of $\mathcal{L}_{\pi}$ unchanged. But this is usually not feasible in symmetric multi-galileon models. For example, in the case of $SO(N)$ fundamental representation, $\pi=(\pi_1,\pi_2,...,\pi_N)$ can not be linked to $\pi'=(\pi'_1,\pi_2,...,\pi_N)$ by an internal $SO(N)$ transformation. (Note that the $SO(N)$ invariant coupling $P(\pi^2)T$, $P(\pi^2)$ being a general function of $\pi^i\pi_i$, has been considered in~\cite{ahkt}, and the authors found gradient instability as well as superluminal excitations for the spherically symmetric background.) We could argue that from the viewpoint of braneworld scenarios the coupling~(\ref{coupling}) (instead of, say, $\pi_1T$) might be what one might expect for symmetric multi-galileon models. In a braneworld setup, the multiple galileon fields living on a brane usually descend from the extra dimension coordinates as functions of the 4D brane volume coordinates~\cite{dbigal,htw}. Since the symmetric multiple galileon fields enjoy some internal symmetry, the extra dimensional coordinates must have the corresponding symmetry at least near the brane. As the near brane geometry is expected to plays a role in determining the coupling to gravity, we may expect the different multiple galileons couple to gravity on a equal or similar basis.

At distances and time scales shorter than the Hubble length, the Friedmann-Robertson-Walker metric can be considered as a perturbation above Minkowski spacetime. Due to the kinetic mixing (\ref{coupling}), the cosmic profile of the multi-galileon can be cast \emph{within the Hubble length} as~\cite{galori}
\be \label{cospi}
\Sigma\pi=-\frac{1}{4}(H^2-H_{\rm gr}^2)x_\mu x^\mu+\frac12(\dot{H}-\dot{H}_{\rm gr})t^2 ,
\ee
where $\Sigma\pi\equiv\pi_1+...+\pi_N$, $H$ is the actual Hubble parameter for a given source $T_{\mu\nu}$ and $H_{\rm gr}$ is the hypothetical Hubble parameter in GR with the same $T_{\mu\nu}$ as the source. Thus the cosmic background configuration of $\Sigma\pi$ is given by $-\frac{1}{4}(H^2-H_{\rm gr}^2)x_\mu x^\mu$. Assuming all the fields have similar coordinate dependence, the vacuum solution is given by
\be 
\bar{\pi}_i=-\frac{1}{4}\bar{k}_i x_{\mu}x^{\mu}, \qquad  \Sigma \bar{k}\equiv \bar{k}_1+...+\bar{k}_N=H^2-H_{\rm gr}^2.
\ee

\section{Goldstone's Theorem in Symmetric Multi-galileon Modified Gravity} \label{goldstonet}

In this section we will see that the symmetric multi-galileon modified gravity exhibits spontaneous breaking of symmetries on some vacuum solution, and for every broken continuous symmetry the theory loses a canonical kinetic term, which resembles the usual Goldstone's theorem in a scalar field theory. We will also discuss the implications of this modified Goldstone's theorem.

\subsection{An Example}

Let us first see a simple example of this theorem: spontaneous breaking of the $SO(N)$ fundamental representation. The most general $SO(N)$ multi-galileon Lagrangian in the fundamental representation is given by~\cite{psz2}
\be \label{sonl}
\mathcal{L}_\pi=-\alpha\,\pd_{\mu}\pi^i\pd^{\mu}\pi_i - \beta\, \delta^{\phantom{[}\rho\mu\lambda}_{[\sigma\nu\tau]}\pd_{\rho}\pi^i\pd^{\sigma}\!\pi_i \pd_{\mu}\pd^{\nu}\!\pi^j \pd_{\lambda} \pd^{\tau}\!\pi_j ,
\ee
where $\delta^{\phantom{[}\rho\mu\lambda}_{[\sigma\nu\tau]}\equiv 3!\delta^{\rho}_{[\sigma}\delta^{\mu}_{\nu}\delta^{\lambda}_{\tau]}$, and $\alpha$ and $\beta$ are free parameters. Varying (\ref{coupling}) with respect to $\pi_i$, we get the equations of motion:
\be \label{eom1}
2 \alpha\,\Box\pi_i + 4 \beta\, \delta^{\phantom{[}\rho\mu\lambda}_{[\sigma\nu\tau]}\pd_{\rho}\pd^{\sigma}\!\pi_i  \pd_{\mu}\pd^{\nu}\!\pi^j \pd_{\lambda} \pd^{\tau}\!\pi_j = - T .
\ee
We would like to see whether there is any \emph{self-accelerating} background (or vacuum) in this theory. By a self-accelerating background, we refer to the case where the universe has a (at least approximately) de Sitter solution without support of a cosmological constant, i.e., the case where $\bar{\pi}_i=-\frac{1}{4}\bar{k}_i x_{\mu}x^{\mu}$ with $\Sigma \bar{k}=H^2>0$ and $T=0$ is a solution to the equations of motion (\ref{eom1}). Substituting this profile into the equations of motion, we get
\be \label{sabg}
-4 \bar{k}_i (\alpha + 3\beta \bar{k}^j \bar{k}_j) = 0  ,
\ee
which reduce to
\be \label{mss}
 \bar{k}_i=0  , 
\ee
or
\be \label{sas}
 \bar{k}^j \bar{k}_j=-\frac{\alpha}{3\beta}  .
\ee
The former solution corresponds to Minkowski spacetime, while the later can be a self-accelerating solution if $\alpha/\beta<0$ and $\Sigma \bar{k}=H^2>0$, which we assume to be satisfied. Note that (\ref{sas}) is not an isolated solution, instead it is a continuum of possible solutions.

Then we would like to see whether the self-accelerating solution can be free of ghosts, negative canonical kinetic terms. To this end, we expand the Lagrangian (\ref{sonl}) above the background (\ref{sas}), i.e., we do the transformation $\pi_i \to \bar{\pi}_i+\pi_i$ and neglect the background part of the Lagrangian:
\begin{align} \label{sonlsa}
\mathcal{L}_{\pi}=&-6\beta\, \pd_{\mu}(\bar{k}^i\pi_i) \pd^{\mu}(\bar{k}^j \pi_j) 
+4\beta\, \bar{k}^i \delta^{\phantom{[}\rho\mu}_{[\sigma\nu]}\pd_{\rho}\pi_i\pd^{\sigma}\!\pi^j \pd_{\mu}\pd^{\nu}\!\pi_j \nn
&-\beta\, \delta^{\phantom{[}\rho\mu\lambda}_{[\sigma\nu\tau]}\pd_{\rho}\pi^i\pd^{\sigma}\!\pi_i \pd_{\mu}\pd^{\nu}\!\pi^j \pd_{\lambda} \pd^{\tau}\!\pi_j  .
\end{align}
Requiring the self-accelerating background to be ghost free gives rise to $\beta>0$, so the conditions for a ghost free self-accelerating solution are
\be
\beta>0~~~\mathrm{and}~~~\alpha<0~~~\mathrm{and}~~~\Sigma \bar{k}=H^2>0 .
\ee
Therefore, in the $SO(N)$ (fundamental) multi-galileon theory, when the self-accelerating branch is ghost free, the Minkowski branch is inevitably haunted by ghosts, and vice versa. Also, we see that there is just one canonical kinetic term on the self-accelerating background, while on the Minkowski background there are $N$ of them\,\footnote{The same result was also reached in~\cite{htw} as we were preparing this paper.}.

All of these would become apparent from a point of view of spontaneous symmetry breaking. To facilitate this approach, we would like to utilize the \emph{action polynomial} introduced in~\cite{psz1}\,\footnote{Note that here we define a slightly different $L(k)$ from that defined in our previous paper. This is because here we write $\mathcal{L}_\pi\sim -\pd \pi \pd \pi \pd\pd\pi...$, while in~\cite{psz1} we use $\mathcal{L}'_\pi\sim \pi\pd \pd \pi \pi \pd\pd\pi...$. These two forms are related by integration by parts in the action, so they are physically equivalent. However, when $\pi$ is evaluated at $\pi=-k_i x_{\mu}x^{\mu}/4$, total derivatives also give rise to terms proportional to $x_{\mu}x^{\mu}$ , so they differ by a factor of $-2$, i.e., $\mathcal{L}'_{\pi}=-2\mathcal{L}_{\pi}$ at $\pi=-k_i x_{\mu}x^{\mu}/4$.}:
\begin{align} \label{actpoly}
L(k)&=-4  \frac{\int\ud^4 x\;\mathcal{L}_\pi}{\int\ud^4 x\; x_\mu x^\mu}\\
      &=\alpha k^i k_i+\frac32 \beta (k^i k_i)^2 ,
\end{align}
where $\pi$ is evaluated at $-k_i x_{\mu}x^{\mu}/4$. By explicit calculation~\cite{psz1}, we have shown that the extrema of $L(k)$ correspond to cosmic background solutions; also, the coefficient matrix of the canonical kinetic terms of the $N$-galileon about a background ($k_i=\bar{k}_i$) is equal to the Hessian of $L(k)$ about the background:
\be \label{cmkinetic}
K_{ij}(\bar{k}) = H_{ij}(\bar{k})\equiv \left. \frac{\pd^2 L(k)}{\pd k^i \pd k^j}\right|_{k=\bar{k}},
\ee
meaning among the extrema only the (local) minima are ghost free ones. These properties of $L(k)$ allow us to treat $L(k)$ as some kind of \emph{effective potential} in finding ghost free vacua. As an aside, note that in canonical field theories such as a scalar field theory, the Hamiltonian of the theory provides an energy function to minimize to find stable vacua. However, due to their non-trivial vacuum configurations and higher derivative nature, the Hamiltonian formulation of multi-galileon theories does not give rise to such a clear energy function for the background configuration $\pi=-k_i x_{\mu}x^{\mu}/4$; see Section~\ref{hamil} for details.

\begin{figure}
\includegraphics[height=2.3in,width=2.3in]{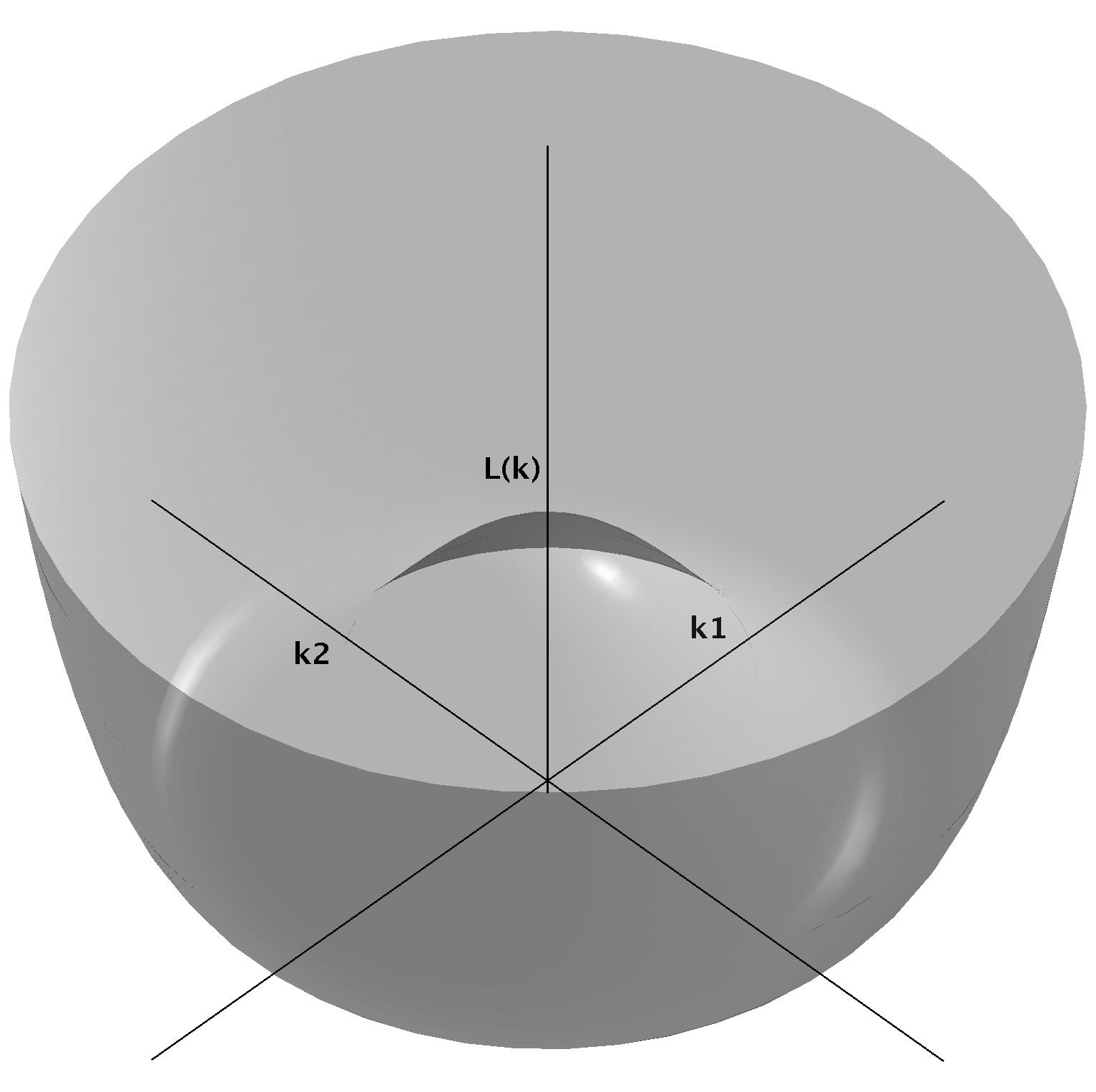}
\caption{The Mexican hat shape of the action polynomial $L(k)$ of the $SO(N)$ multi-galileon (plotted for the case of $SO(2)$). The vacuum rests on the trough, spontaneously breaking $SO(N)$ to $SO(N\!-\!1)$, therefore fluctuations along the $N-1$ flat directions of the trough do not have canonical kinetic terms.} \label{mexhat}
\end{figure}

Now, we can easily recover the results of the $SO(N)$ multi-galileon vacuum solutions using $L(k)$. The extrema of $L(k)$ give rise to the Minkowski background $\bar{k}_i=0$ and the self-accelerating background $\bar{k}^j \bar{k}_j=-\alpha/3\beta$. The background $\bar{k}^j \bar{k}_j=-\alpha/3\beta$ is a minimum of $L(k)$ only if $\alpha<0$ and $\beta>0$. Also, since the continuum $\bar{k}^j \bar{k}_j=-\alpha/3\beta$ is a minimum, topologically $\bar{k}_i=0$ can not be a minimum, thus for the same set of parameters only one of the two backgrounds can be stable. The Hessian of $L(k)$ about the self-accelerating background is given by $K_{ij}(\bar{k})=12\beta \bar{k}_i \bar{k}_j$, which has only one non-zero eigenvalue, so there is just one canonical kinetic term on this background. Indeed, we might visualize $L(k)$ with a ghost free self-accelerating background by a ``Mexican hat'' (Fig.~\ref{mexhat}). The trough of this Mexican hat is an $(N\!-\!1)$-sphere, respecting $SO(N)$. An $(N\!-\!1)$-sphere (or $SO(N)$) has $N(N-1)/2$ independent rotational symmetries. The vacuum solution occupies one point on the trough and thus only respects an $SO(N\!-\!1)$ subgroup, which leaves a sub $(N\!-\!2)$-sphere still rotational symmetric and breaks $N-1$ rotational symmetries. Only the radial direction around the trough accommodates non-trivial ``oscillations'', reflecting the presence of only one canonical kinetic term. The $N-1$ flat directions represent the loss of $N-1$ canonical kinetic terms.

\subsection{General Proof}

This is of course reminiscent of Goldstone's theorem for a canonical scalar field theory with a potential. Here we are able to prove an analogous theorem for a symmetric multi-galileon theory with an arbitrary internal group that the number of canonical kinetic terms that are lost is equal to the number of spontaneously broken symmetries, which in turn equals the dimension of the total symmetry group minus that of the unbroken subgroup. Again it is sufficient to use the action polynomial $L(k)$ to prove this.

Let $k=\bar{k}_i$ be a (local) minimum of $L(k)$, so it is a sensible background to expand the theory. Since $k=\bar{k}_i$ is a minimum, $K_{ij}(\bar{k}_i)$ should only have non-negative eigenvalues. The eigenvectors of positive eigenvalues correspond to the canonical kinetic terms, while the eigenvectors of zero eigenvalues correspond to the degrees of freedom without canonical kinetic terms.

To prove the theorem, we must show that every spontaneously broken symmetry gives rise to an independent zero-eigenvalued eigenvector. Under an infinitesimal group action, for the configuration $\pi^i=-k^i x_{\mu}x^{\mu}/4$, we have
\be    \label{gptf}
k^i \to  k^i+\epsilon \;\Delta^i(k)  ,
\ee
where $\epsilon$ is an infinitesimal. Since $\mathcal{L}_\pi$ is invariant under a group transformation, from (\ref{actpoly}), we infer that $L(k)$ is also invariant. So we have
\be
L(k)=L(k+\epsilon \; \Delta(k))=L(k)+\epsilon \; \frac{\pd L(k)}{\pd k^i}\Delta^i(k) ,
\ee
which leads to the identity
\be
\frac{\pd L(k)}{\pd k^i}\Delta^i(k)=0  .
\ee
Differentiating it with respect to $k^i$ and evaluating it at the vacuum of the theory ($k_i=\bar{k}_i$) gives 
\be \label{cmdrel}
K_{ij}(\bar{k})\Delta^j(\bar{k})=0  ,
\ee
where $K_{ij}(\bar{k})$ is the coefficient matrix of the canonical kinetic terms, as defined in (\ref{cmkinetic}). Now, if the transformation (\ref{gptf}) belongs to the unbroken subgroup, the vacuum $k_i=\bar{k}_i$ is invariant under the transformation and the relation (\ref{cmdrel}) is trivial as we have $\Delta^j(\bar{k})=0$. If the transformation (\ref{gptf}) belongs to a spontaneously broken symmetry, the vacuum is changed along the flat directions of the continuous minimum of $L(k)$ and so we have  $\Delta^j(\bar{k})\neq 0$. In this case, $K_{ij}(\bar{k})$ has a zero eigenvalue and the eigenvector $\Delta^j(\bar{k})$, or $\Delta^j(\pi)$, is the degree of freedom that loses its canonical kinetic term.

\subsection{Implications}

In multi-galileon theories, due to the presence of higher order kinetic terms, absence of a canonical kinetic term does not necessarily mean loss of a dynamical degree of freedom. Taking the $SO(N)$ multi-galileon theory for example, by integration by parts the cubic term of the Lagrangian above the self-accelerating background (\ref{sonlsa}) can be cast as
\begin{align}
\mathcal{L}_{\pi}^{(3)} =-4\beta \pd^a\pd_a (\bar{k}^j \pi_j)\, \dot{\pi}^i \dot{\pi}_i 
      -8\beta \pd^a\pd_a (\bar{k}^j \pi_i)\, \dot{\pi}^i \dot{\pi}_j+  4\beta\, \delta^{\phantom{[}ac}_{[bd]}\pd_{a} \pi^i \pd^{b}\!\pi_i\, \pd_{c} \pd^{d}(\bar{k}^j \pi_j)   ,
\end{align}
where $a,b,c,d$ are spatial indices (rather than group indices), and the theory has $N$ cubic kinetic terms. The conjugate momenta of $\dot{\pi}^i(\bfx,t)$ is non-vanishing and the canonical phase space is non-trivial for all the $N$ degrees of freedom. So there are still $N$ apparent dynamical degrees of freedom on the self-accelerating background.

However, since a mode without a canonical kinetic term can be regarded as infinitely strongly coupled and Vainshtein mechanism takes effect in galileon models, these modes would self-screen themselves from the others. We shall demonstrate this schematically. Suppose $\pi_1$ loses its canonical kinetic term around the self-accelerating vacuum and consider a slightly different background where the Lagrangian is given by
\be
\mathcal{L}\sim \epsilon^2 M_P^2 \pd \pi_1 \pd \pi_1 + \frac{M_P^2}{M^2} \pd \pi_1 \pd \pi_1 \pd \pd \pi_1 + \pi_1 T+ ... \,,
\ee 
with $...$ standing for other interactions and modes. To see the genuine dynamics of this mode, we canonically normalizing it, which gives rise to
\be
\mathcal{L}\sim  M_P^2 \pd \tilde{\pi}_1 \pd \tilde{\pi}_1 + \frac{1}{\epsilon^3}\frac{M_P^2}{M^2} \pd \tilde{\pi}_1 \pd \tilde{\pi}_1 \pd \pd \tilde{\pi}_1 + \frac1{\epsilon}\tilde{\pi}_1 T  + ... \,.
\ee
We can recover the perturbative Lagrangian around the vacuum by taking the limit $\epsilon\to 0$, where we can clearly see that $\pi_1$ is infinitely strongly coupled. Now, we can calculate that the Vainshtein radius of a spherical source ($M_{\rm s}$) for $\pi_1$ (see e.g.~\cite{psz1}):
\be
R_V=\left(\frac1{\epsilon}\right)^{\frac43}\left(\frac{M_{\rm s}}{M_P^2M^2}\right)^{\frac{1}{3}} .
\ee 
It goes to infinity when $\epsilon$ goes to 0, meaning this mode would be self-screened at infinitely large distances, and thus is effectively non-dynamical on the vacuum, at least in terms of weak gravitational interactions. Nevertheless, as mentioned above, although some modes in the galileon multiplet lose their canonical kinetic terms on a self-accelerating vacuum, these modes can re-acquire their quadratic kinetic terms on backgrounds with matter sources. Therefore, around a generic background such as in the solar system, these modes are indeed not strongly coupled. As an aside, if there is a cosmological constant, the multi-galileon internal symmetry will be explicitly broken, in which case there is generally no loss of canonical kinetic terms. For the example of $SO(2)$ fundamental representation, the action polynomial is deformed to be a tilted Mexican hat where there is only an unique minimum.

When calculating the leading corrections to GR, thanks to Vainshtein effect, we might simply exclude these inert modes. So the spontaneous symmetry breaking and the subsequent freeze-out of some dynamical modes could be reflected in tests of modification to gravity force, as the leading corrections are encoded in the canonical kinetic terms. We still take the $SO(N)$ multi-galileon for example. First, note that the one particle exchange amplitude between two conserved sources $T_{\mu\nu}$ and $T'_{\mu\nu}$ in GR schematically is given by
\be
\mathcal{A}_{GR}\sim -\frac2{M_P^2}\left(T^{\mu\nu}\frac{1}{\Box}T'_{\mu\nu}-\frac12 T\frac{1}{\Box}T'\right) .
\ee
For simplicity, we assuming $\bar{k}_i\sim \bar{k}$, so we have $N\bar{k}^2\sim -\alpha/3\beta$. When the vacuum is spontaneously broken and rests on the self-accelerating branch (\ref{sas}), from (\ref{sonlsa}) we can see that the $SO(N)$ multi-galileon gives rise to a leading correction
\be \label{modsa}
\delta \mathcal{A}_{SA}=\mathcal{A}-\mathcal{A}_{GR}\sim \frac{N}{2\alpha } T\frac{1}{\Box}T'.
\ee
This is compared to the case without spontaneous symmetry breaking, when the leading correction on the Minkowski branch (\ref{mss}) is given by
\be
\delta \mathcal{A}_{M}\sim -\frac{N}{\alpha} T\frac{1}{\Box}T'  .
\ee

On the other hand, when testing the multi-galileon modification to gravitational force upto leading order, we have to deal with observational degeneracy between multi-galileon theories with different internal symmetries and different choices of vacuum branches. Again taking the $SO(N)$ example and assuming $\bar{k}_i\sim \bar{k}$, the leading correction from the $SO(N)$ multi-galileon on the self-accelerating branch (\ref{modsa}) is the same as that from a multi-galileon theory without internal symmetries and with canonical kinetic terms $-\alpha(\pd\pi_1\pd\pi_1+...+\pd\pi_{N/2}\pd\pi_{N/2})$, provided $N$ is an even number.

\section{Hamiltonian Formulation of Multi-galileon Theories} \label{hamil}

In this section, we deviate from our main plot of the paper and briefly introduce a subplot: the Hamiltonian approach of multi-galileon theories. First, we derive the Hamiltonian for a general multi-galileon theory with or without internal symmetries.

As the Lagrangian of a multi-galileon theory contains terms with more than 2 spacetime derivatives, one might expect the Hamiltonian formulation of a multi-galileon theory should involve Ostrogradski's prescription for high order derivative theories (see for example~\cite{ostrog}). However, a bell should be certainly rung to this naive thinking once we notice the fact that the equations of motion of a multi-galileon theory has only second order derivatives. We will see that a general multi-galileon Lagrangian can be cast to have only up to first order time derivatives. A general multi-galileon theory without a tadpole term can be written as~\cite{psz2}
\begin{align} \label{laggel}
\hat{\mathcal{L}}_\pi =-\sum_{n=2}^5 \;&\alpha^{i_{ 1}...i_{ n}}\,\delta^{\phantom{[}\mu_{ 2}...\mu_{ n}\phantom{]}}_{[\nu_{ 2}...\nu_{ n}]}
\pd_{\mu_{2}}\pi_{i_{ 1}}\pd^{\nu_{ 2}} \pi_{i_{ 2}}
\pd_{\mu_{ 3}}\pd^{\nu_{ 3}} \pi_{i_{ 3}}...\pd_{\mu_{ n}}\pd^{\nu_{ n}} \pi_{i_{ n}},
\end{align}
where $\delta^{\,\mu_{ 2}...\mu_{ n}}_{[\nu_{ 2}...\nu_{ n}]}\equiv (n-1)!\delta^{\:\mu_2}_{[\nu_2}... \delta^{\mu_n}_{\nu_{ n}]}$, $i_1,...,i_n$ label different galileons (not necessarily internal group indices) and summation over repeated $i_k$ is understood. $\alpha^{i_1 ... i_n}$ are free parameters of the theory, and can be chosen as symmetric in exchanging the indices since $\delta^{\phantom{[}\mu_{ 2}...\mu_{ n}\phantom{]}}_{[\nu_{ 2}...\nu_{ n}]}
\pd_{\mu_{2}}\pi_{i_{ 1}}\pd^{\nu_{ 2}} \pi_{i_{ 2}}
\pd_{\mu_{ 3}}\pd^{\nu_{ 3}} \pi_{i_{ 3}}...\pd_{\mu_{ n}}\pd^{\nu_{ n}} \pi_{i_{ n}}$ can be made symmetric in exchanging the galileon indices by integration by parts.

To see what the derivative structure is, we should unfold the anti-symmetrisation. Since the Hamiltonian formulation only requires the knowledge of the time derivative structure, we only need to separate the time derivatives from the spatial ones. A useful relation for the separation is 
\be
\delta^{\phantom{[}\mu_{ 2}...\mu_{ n}\phantom{]}}_{[\nu_{ 2}...\nu_{ n}]}T^{\nu_2...\nu_n}_{\mu_2...\mu_n}=  \delta^{\phantom{[}a_{ 2}...a_{ n}\phantom{]}}_{[b_{ 2}...b_{ n}]}T^{b_2...b_n}_{a_2...a_n} + \sum_{i=2}^n\sum_{j=2}^n\left.\delta^{\phantom{[}a_{ 2}...a_{ n}\phantom{]}}_{[b_{ 2}...b_{ n}]}T^{b_2...b_n}_{a_2...a_n}\right|_{\!\scriptsize \begin{array}{c} a_i\! \to \!t_1 \\ b_j\! \to \!t_2 \end{array}} ,
\ee  
where $T^{\nu_2...\nu_n}_{\mu_2...\mu_n}$ is an arbitrary tensor, $t_1$ and $t_2$ are time indices, and $a_i$ and $b_i$ are spatial indices. The double summation is over replacement of one up spatial index with $t_1$ and one down spatial index with $t_2$, so there are $(n-1)^2$ terms with time derivatives. Applying this formula to (\ref{laggel}) and repeatedly integrating by parts, we can see that for $n$-th order a term with $\delta^{...t_1 a_i...}_{...t_2 b_i...}$ gives rise to $\alpha^{i_{ 1}...i_{n}} \delta^{\phantom{[} a_3...a_{ n}\phantom{]}}_{[ b_3...b_{ n}]}\dot{\pi}_{i_1} \!\dot{\pi}_{i_2}  \pd_{a_{ 3}}\!\pd^{b_{ 3}} \pi_{i_{ 3}}...\pd_{a_{ n}}\!\pd^{b_{ n}} \pi_{i_n}$, while a term with $\delta^{...t_1 a_i...}_{... b_i t_2...}$ only gives rise to half of that, with all the other terms cancelling each other. Therefore the Lagrangian (\ref{laggel}) can be cast as
\begin{align} \label{gallagh}
&\hat{\mathcal{L}}_\pi =  \sum_{n=2}^5  \,  \alpha^{i_{ 1}...i_{n}} \left[C^2_n  \delta^{\phantom{[} a_3...a_{ n}\phantom{]}}_{[ b_3...b_{ n}]}\dot{\pi}_{i_1} \!\dot{\pi}_{i_2} \!- \delta^{\phantom{[}a_{ 2}...a_{ n}\phantom{]}}_{[b_{ 2}...b_{ n}]}
\pd_{a_{2}}\pi_{i_{ 1}}\pd^{b_{ 2}} \pi_{i_{ 2}} \right] \pd_{a_{ 3}}\!\pd^{b_{ 3}} \pi_{i_{ 3}}...\pd_{a_{ n}}\!\pd^{b_{ n}} \pi_{i_n} ,
\end{align}
where $C^2_n\equiv n(n-1)/2$. The appearance of the combinatorial number $C^2_n$ is what one might expect, since the indices $i_1,...,i_n$ are symmetric and so there are $C^2_n$ ways to pick out two $\pi_i$s with first order time derivatives. Due to the first order structure in time derivatives, we can simply take $\pi_i({\bf x},t)$ canonical coordinates and define the conjugate momenta as 
\begin{align} \label{phidef}
\phi^i({\bf x},t) &  = \frac{\pd \hat{\cal L}_{\pi}}{\pd \dot{\pi}_i({\bf x},t)} \nn
& = 2  \sum_{n=2}^5     \alpha^{i i_2...i_{n}} C^2_n \delta^{\phantom{[} a_3 ...a_{ n}\phantom{]}}_{[b_3...b_{ n}]}\dot{\pi}_{i_2}\pd_{a_{ 3}}\!\pd^{b_{ 3}} \pi_{i_3}...\pd_{a_{ n}}\!\pd^{b_{ n}} \pi_{i_n} .
\end{align}
Defining the matrix
\be
M^{ij}\equiv 2 \sum_{n=2}^5   \alpha^{i j...i_{n}} C^2_n \delta^{\phantom{[}a_3 ...a_{ n}\phantom{]}}_{[b_3...b_{ n}]}\pd_{a_{ 3}}\!\pd^{b_{ 3}} \pi_{i_3}...\pd_{a_{ n}}\!\pd^{b_{ n}} \pi_{i_n} , 
\ee
we can reverse (\ref{phidef}) and get
\be
\dot{\pi}_i=(M^{-\!1})_{ij}\phi^j  .
\ee
To get the Hamiltonian of the multi-galileon theory, we perform the Legendre transformation
\begin{align} \label{hamilt}
\hat{H}_{\pi} =\int\! \ud^3 x \left[ \dot{\pi}_i\phi^i-\hat{\mathcal{L}}_{\pi}  \right]  = \int\! \ud^3 x\, \hat{\mathcal{H}}_{\pi}  ,
\end{align}
where the Hamiltonian density is given by
\begin{align}  \label{hamdensity}
\hat{\mathcal{H}}_\pi &=  \sum_{n=2}^5  \,  \alpha^{i_{ 1}...i_{n}}  \left[C^2_n \delta^{\phantom{[} a_3...a_{ n}\phantom{]}}_{[ b_3...b_{ n}]}\dot{\pi}_{i_1} \!\dot{\pi}_{i_2} \!+ \delta^{\phantom{[}a_{ 2}...a_{ n}\phantom{]}}_{[b_{ 2}...b_{ n}]}
\pd_{a_{2}}\!\pi_{i_{ 1}}\!\pd^{b_{ 2}} \pi_{i_{ 2}} \right] \pd_{a_{ 3}}\!\pd^{b_{ 3}} \pi_{i_{ 3}}...\pd_{a_{ n}}\!\pd^{b_{ n}} \pi_{i_n} \nn
 & = \frac12 (M^{-\!1})_{ij}\phi^i\phi^j + \sum_{n=2}^5 \alpha^{i_{ 1}...i_{n}}  \delta^{\phantom{[}a_{ 2}...a_{ n}\phantom{]}}_{[b_{ 2}...b_{ n}]}
\pd_{a_{2}}\!\pi_{i_{ 1}}\!\pd^{b_{ 2}} \pi_{i_{ 2}}  \pd_{a_{ 3}}\!\pd^{b_{ 3}} \pi_{i_{ 3}}...\pd_{a_{ n}}\!\pd^{b_{ n}} \pi_{i_n}   .
\end{align}

Now, we would like to know what the Hamiltonian looks like for the vacuum configuration $\pi_i=-k_i x_{\mu}x^{\mu}/4$:
\begin{align} \label{hamfun}
H_{\pi} = \frac14 \int\! \ud^3 x \,&\left[ (t^2+{\bf x}^2) L^{(2)}(k)
+ (3t^2+\frac23 {\bf x}^2)  L^{(3)}(k)
+ (6t^2+\frac13 {\bf x}^2)  L^{(4)}(k)\right.   \nn
&\left. ~ + (10t^2) \, L^{(5)}(k)  \phantom{\frac11} \!\!\!\!\right]  ,
\end{align}
where $L^{(i)}(k)$ are the $i$-th order terms of the action polynomial $L(k)$. In a canonical field theory with a constant field background, since the Hamiltonian is an (infinitely) extensive quantity, we can divide the Hamiltonian by the volume of the spacetime to extract an energy function of the constant field, which can be minimized to find the vacua of the theory. Here we find the same procedure is not applicable to a multi-galileon modified gravity theory, as we can see from (\ref{hamfun}) that the ``volume factor'' is different for different orders of $k_i$. This of course originates from the high derivative nature of multi-galileon theories and the non-trivial background configuration $\pi_i=-k_i x_{\mu}x^{\mu}/4$. Note that for a multi-galileon action with the configuration $\pi_i=-k_i x_{\mu}x^{\mu}/4$, a total derivative (say, $\pd_{t}(\pi_1\pd^{t}\pi_2\pd_{i}\pd^{i} \pi_3 )$) will actually give rise to nontrivial contribution ($-3k_1 k_2 k_3 (3t^2-\bfx^2)/16$). Indeed, from (\ref{laggel}) to (\ref{gallagh}) we have performed a series of integration by parts and neglected the subsequent total derivatives, which is responsible for the different ``volume factors'' in (\ref{hamfun}).

We also note that the Hamiltonian density (\ref{hamdensity}) (hence the Hamiltonian) is generally unbounded from below, i.e., the Hamiltonian density can be arbitrarily lowered by choosing suitable initial field configurations. This is due to the presence of higher than quadratic order multi-galileon terms and because there are terms where the first derivatives of galileon fields are not in ``squared'' forms (e.g., $\dot{\pi}_1 \dot{\pi}_1 (\pd_a \pd^a \pi_2)^2$ is ``squared'', but $\pd_a \pi_1  \pd^a \pd_b \pi_1 \pd^b \pi_1 \pd^{c} \pd_c \pi_1$ and $\dot{\pi}_1 \dot{\pi}_2 \pd_a \pd^a \pi_3$ are not.). Since we know the galileon models define a conventional Cauchy problem, the galileon fields and their first derivatives can be arbitrarily chosen. By making the first derivatives of galileons increasingly steep, we can lower the Hamiltonian density arbitrarily. Note that, even for the background configuration $\pi_i=-k_i x_{\mu}x^{\mu}/4$, the Hamiltonian density at a fixed spacetime point is not bounded below if the highest order of galileon terms is odd. The perturbative Hamiltonian above some self-accelerating background ($k_i=\bar{k}_i$) can also be cast in the form (\ref{hamilt}) with the parameters $\alpha^{i_{ 1}...i_{n}}$ replaced by a new set of parameters $\beta^{i_{ 1}...i_{n}}(\bar{k})$ (as polynomials of $\bar{k}_i$) (see e.g.~\cite{psz1}), so it is also unbounded below. In a fundamental theory, this of course signals instabilities. However, the multi-galileon modified gravity is only supposed to be the decoupling limit of some underlying full theory, so one should really check whether the Hamiltonian of the underlying full theory is well behaved or not. The underlying theory presumably has 4D diffeomorphism invariance, so the corresponding naive 4D Hamiltonian (excluding the part from extra dimensions) is tuned to zero by 4 constraint equations, similar to that in GR. A useful 4D Hamiltonian arises when the theory is ``deparameterized''~\cite{adm}, but from the experiences in GR, even checking the positivity of the background solution could be nontrivial\,\footnote{Nevertheless, for the case where the multi-galileon Hamiltonian density (\ref{hamilt}) is included in the constraint equations, the multi-galileon Hamiltonian (\ref{hamilt}) being unbounded below is irrelevant to the stability issue of the full theory.}.

On the other hand, due to the derivative structure of the multi-galileon theories, the most negative Hamiltonian value is achieved by setting the gradients close to the cutoff of the theory, i.e., $\pd\sim \Lambda_{\rm cutoff}$. This kind of being unbounded below pushes the limit of a classical theory, as it relies on a small region of the canonical phase space, so one might also doubt whether quantum corrections can alther the picture. A famous example of this is the Hydrogen atom: The classical Coulomb potential for this system ($-e^2/r$) can be made arbitrarily negative by placing the electron close to the nucleus, but the Hydrogen atom is stable upon quantisation of electrodynamics.

\section{Conclusion}

We have coupled the multi-galileon theory with internal symmetries studied in~\cite{psz2} to conventional General Relativity (GR) and proposed it as a modified gravity theory in the decoupling limit where the multi-galileon modifies GR only by mixing with the transverse graviton. We have discussed the phenomenon of spontaneous symmetry breaking of these theories on (classical) self-accelerating vacua. We point out that, similar to that in canonical scalar field theories, the pattern of the symmetry breaking is governed by a new version of Goldstone's theorem that for every broken continuous symmetry the theory loses a canonical kinetic term. Note that as the energy-nomentum tensor $T^{\mu\nu}$ by definition vanishes in the self-accelerating vacuum, this theorem is largely insensitive to the coupling to GR. But we do assume the background configuration of the multi-galileon is given by $\pi_i=-k_i x^{\mu} x_{\mu}/4$. We have also discussed implications of this theorem. In particular, we suggest that the mode that loses its canonical kinetic term, although apparently non-trivial in the phase space, becomes inert due to Vainshtein mechanism. This would lead to different modification to gravitational force, compared to what one would naively expect from the Lagragian with the broken vacuum hidden. Also, there would be degeneracy among multi-galileon theories with different internal symmetries and different choices of vacuum branches.

We have also derived the Hamiltonian of a general multi-galileon theory. We find the Hamiltonian with the configuration $\pi_i=-k_ix^{\mu}x_{\mu}/4$ does not give rise to a good ``effective potential'' to minimize to find the background solution. Besides, we find the Hamiltonian is not bounded below because of the higher order multi-galileon terms. We speculate this pathology might arise from the decoupling limit or the classical nature of multi-galileon theories and argue that the underlying full theory for the multi-galileon or even its quantum version should be investigated to decide whether this is a real problem or not. There are a few attempts to put the galileon description in a more formal framework~\cite{galori,covgal,dbigal,htw}, and it is interesting to see whether the vacuum Hamiltonian in these models is bounded below, which we leave for future work.

~\\

{\bf Acknowledgements}:~I would like to thank Ed Copeland, Antonio Padilla and Paul Saffin for helpful discussions. I also thank Paul Saffin and Antonio Padilla for reading through the manuscript and making valuable suggestions.

\end{document}